\documentclass[twocolumn,prl,superscriptaddress,showpacs,byrevtex]{revtex4}
\usepackage{mathptmx,amssymb}
\usepackage{amsmath}
\usepackage{graphicx}
\usepackage{color}
\usepackage{epstopdf}

\begin{document}

\title{Decay rates of magnetic modes below the threshold of a turbulent dynamo}

\author{J. Herault, F. P\'etr\'elis, S. Fauve}
\affiliation{Laboratoire de Physique Statistique, Ecole Normale Sup\'erieure, CNRS, Universit\'e P. et M. Curie, Universit\'e Paris Diderot, Paris, France}

\begin{abstract} 
We measure the decay rates of magnetic field modes in a turbulent flow of liquid sodium below the dynamo threshold. We observe that turbulent fluctuations induce energy transfers 
between modes with different symmetries (dipolar and quadrupolar).  
Using symmetry properties, we show how to measure the decay rate of each mode without being restricted to 
the one with the smallest damping rate. We observe that the respective values of the decay rates of these modes depend on the shape of the propellers driving the flow. Dynamical regimes, including field reversals,
are observed only when the modes are both nearly marginal. This is in  line with a recently proposed model. 
\end{abstract}
\maketitle

The generation of magnetic field by the flow of an electrically conducting fluid through the dynamo process has been primarily studied in order to understand planetary and stellar magnetic fields \cite{cosmic}.
This phenomenon also provides a canonical example of an instability that occurs on a fully turbulent flow. Instabilities occurring on turbulent flows involve several fundamental open questions, such as the characterization of growing or decaying eigenmodes in the linear regime or the possibly anomalous scaling of the moments of the magnetic field amplitude versus the distance to instability threshold in the nonlinear regime \cite{petrelis2007,exposant}. We present here an experimental study related to the former aspect. 

The threshold of a linear instability of a stationary or time-periodic state is reached when the largest growth rate (eigenvalue) of the eigenmodes  of the evolution equation for the perturbations vanishes. Below this threshold, small perturbations decay exponentially with a decay rate inversely proportional to the corresponding eigenvalue. The value of the decay rate thus measures the distance to the threshold of linear instability of the corresponding eigenmode. Measurements of growth and decay rates are commonly performed to characterize instabilities. Eigenmodes with different symmetries are not linearly coupled and depending on the symmetries of the initial conditions, their decay rate can be measured. The problem is more complex for an instability generated by a turbulent flow. First, although growth and decay can be characterized by a Lyapunov exponent, eigenmodes cannot be defined in the usual way. In the limit of small turbulent fluctuations about the mean flow, one can consider the eigenmodes of the mean flow as a first approximation as done in the past for the dynamo problem \cite{tilgner2002}. However, it has been shown both experimentally \cite{P3} as well as from numerical simulations \cite{gissinger2009} that the mean magnetic field generated by a turbulent flow can strongly differ from a prediction based on the mean flow alone. In addition, even if some modes are defined in an appropriate way, turbulent fluctuations can transfer energy from one mode to the other, thus contaminating decay rate measurements. We report an experimental procedure to define these modes and avoid this bias by independently measuring  the decay rate of the two less damped modes with different symmetries.  We then show how their difference in decay rates can be related to the dynamics observed above instability threshold.

\begin{figure}[htb!]
\begin{center}
\includegraphics[width=60mm,height=60mm]{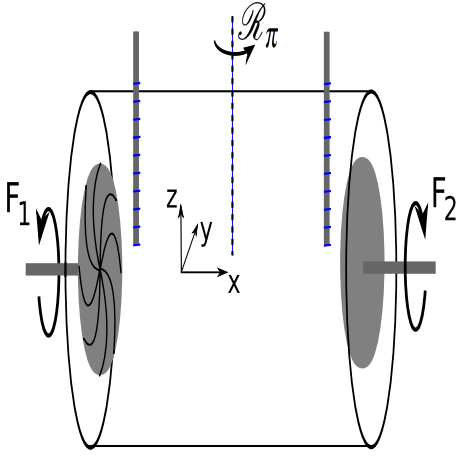}
\caption{Sketch of the experiment: two impellers counter-rotate at frequencies $F_1$ and $F_2$. When $F_1=F_2$ the setup is invariant to $\mathcal{R}_\pi$ rotation. The vertical thick lines indicate the locations of the two arrays of $3$-axis Hall probes which measure the magnetic field close to each disk.}
\label{fig1}
\end{center}
\vspace{- 1cm}
\end{figure}

The experiment concerns the generation of a magnetic field by a von Karman flow of liquid sodium (VKS experiment) that has been already reported in detail elsewhere \cite{Monchaux2009}. The flow is driven by two counter-rotating coaxial propellers in a cylinder containing roughly  $160$ liters of liquid sodium maintained at a temperature around $120^o$ C for which the electrical conductivity $\sigma$ is close to $10^7$ $(\Omega m)^{-1}$ (see fig. \ref{fig1}). The propellers are soft iron disks fitted with curved (resp. straight) iron blades. When they counter-rotate with the same speed, the dynamo threshold is reached for $F_1=F_2=13$ Hz (resp. $19$ Hz). This corresponds to a magnetic Reynolds number, $R_m=2 \pi \mu_0 \sigma R^2 F$ around $24$ (resp. 35) where  $R=154.5$ mm is the radius of the disks. The time-averaged magnetic field is roughly an axial dipole which amplitude displays a slightly imperfect bifurcation because of the remanent magnetization of the disks. When the disks counter-rotate at  different frequencies,  a mode of quadrupolar symmetry is also generated \cite{P3}. 
The dipole and the quadrupole have opposite symmetries with respect to a rotation of $\pi$ around the $z$-axis in the mid-plane between the disks  ($\mathcal{R}_\pi$ as sketched in fig. \ref{fig1}). For a large enough frequency difference, dynamical regimes can be observed such as periodic or random reversals of the magnetic field \cite{P2,P3} that involve energy transfers between the dipole and the quadrupole \cite{gissinger}.

From now on, we restrict to measurements performed for exact couter-rotation $F_1=F_2=F$.  Below the onset of instability, we apply a magnetic field generated by two axial coils. Each coil carries an equal electric current flowing in the same (resp. opposite) direction, thus providing an Helmholtz (resp. anti-Helmholtz) configuration generating an axial field of dipolar (resp. quadrupolar) symmetry, {\it i.e.} odd (resp. even) under $\mathcal{R}_\pi$. 
The magnetic field is measured with two probes located close to the disks, $109$ mm away from the midplane. Each probe measures the three components of the magnetic field at ten positions (the deepest probe is $103$ mm away from the cylinder axis and the distance between probes is $28$ mm).

\begin{figure}[htb!]
\begin{center}

\includegraphics[width=70mm,height=70mm]{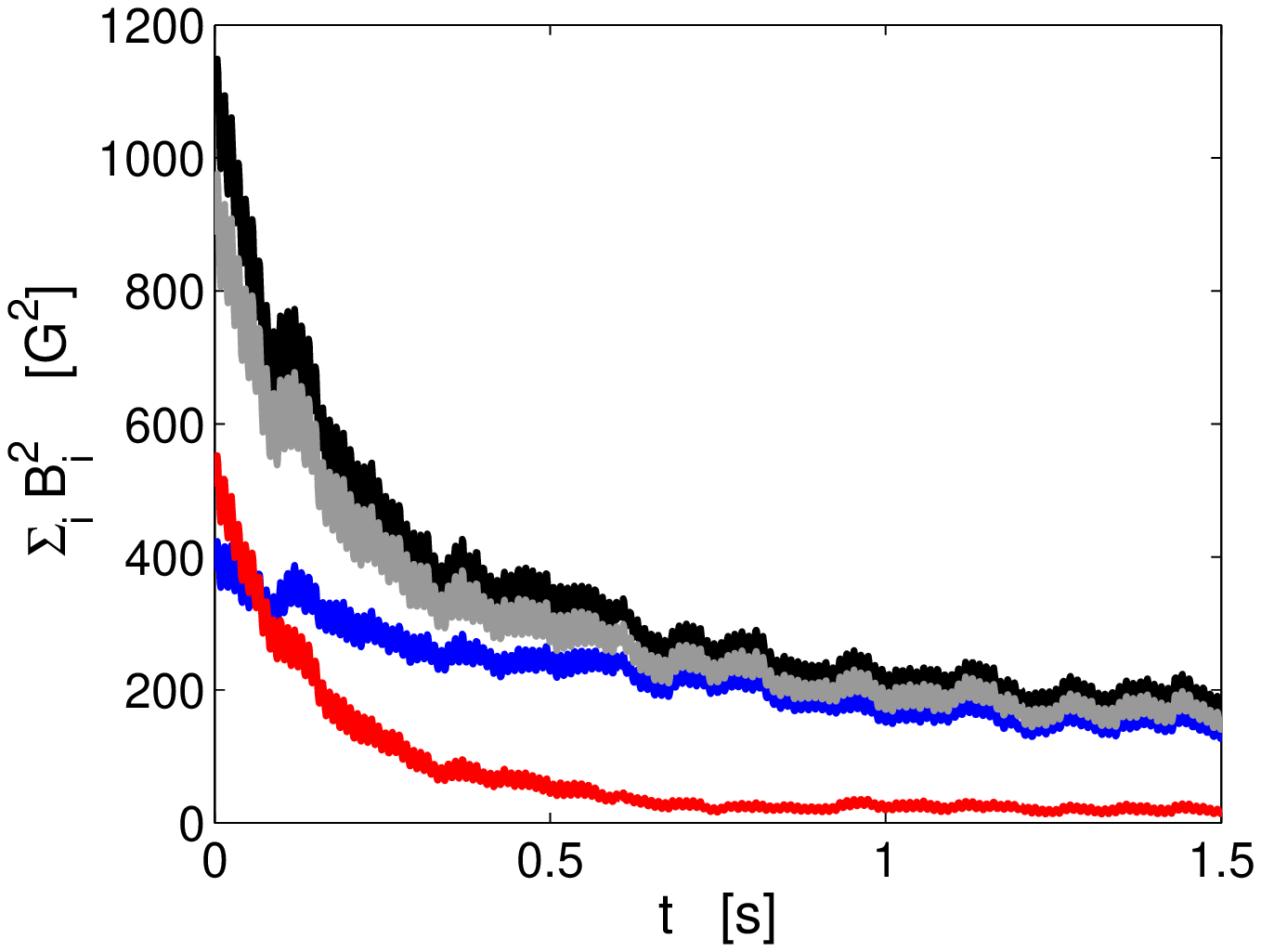}
\includegraphics[width=70mm,height=70mm]{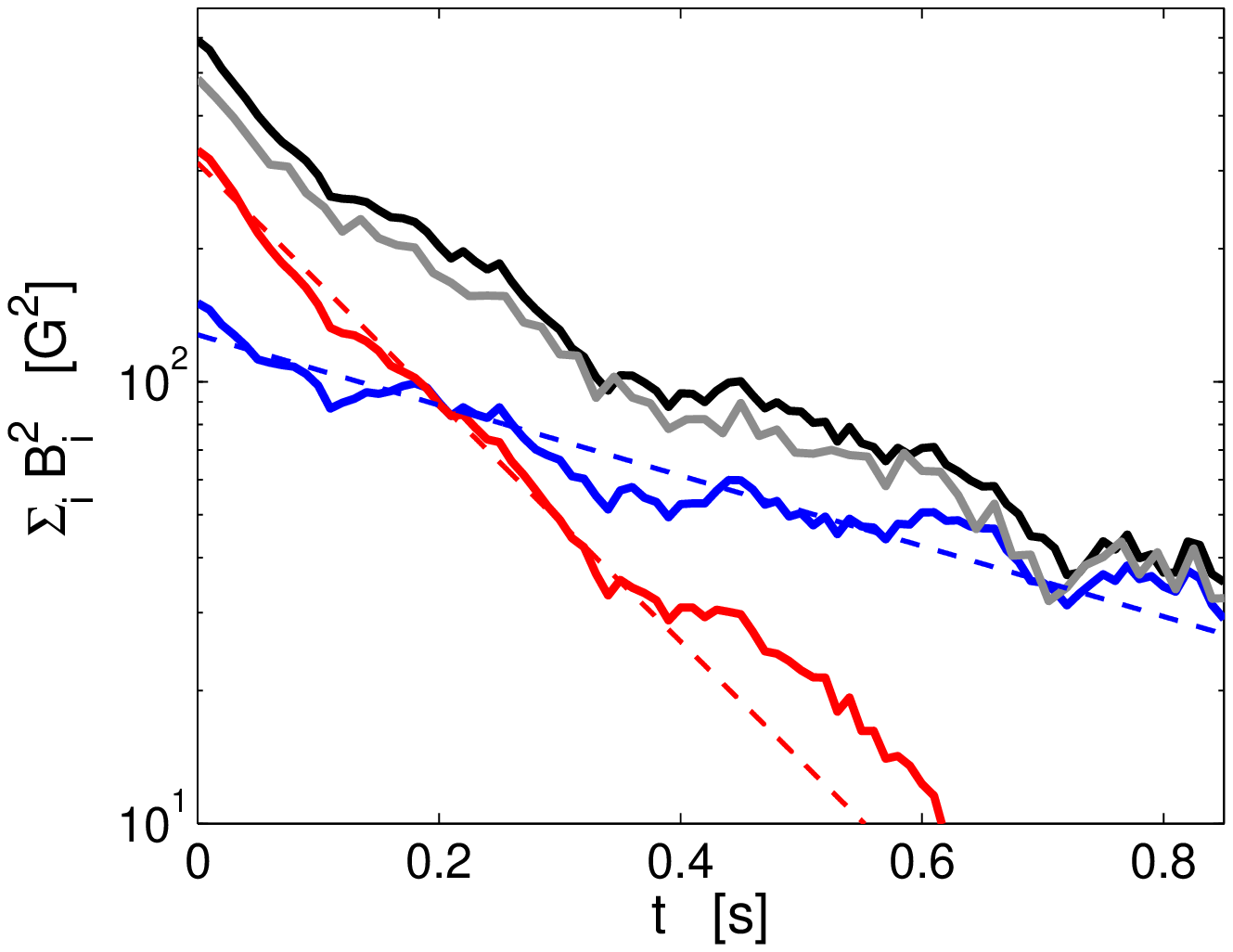}
\caption{
For an anti-Helmotz applied magnetic field, time series of  $\langle \vert B_1\vert^2+\vert B_2 \vert^2 \rangle /2 $ (in black). Two behaviors are
observed. The decay rate measured at intial time  is the same as the one
obtained from the quadrupolar projection (in red). At long time, it is the
same as the one obtained from the dipolar projection (in blue). The grey
curve represents the total amount of energy in the quadrupolar and dipolar
modes.  Top: lin-lin scale; bottom: log-lin scale, the long time value is substracted so that exponential decays appear as straight lines.}
\label{fig2}
\end{center}
\end{figure}

We apply a time-periodic magnetic field with a square wave shape in order to measure the decay rate of the magnetic response below the dynamo threshold as performed in \cite{lathrop,P10}. Experiments typically involve $20$ periods of duration $10\, s$. When  the external magnetic field is on, we observe a field induced by the flow of liquid sodium. When the external field is switched off, the induced magnetic field decays and reaches a small value that results from the ambient field.  A first way to evaluate the amplitude of the excited magnetic  mode is to calculate its energy density defined as the sum over the probes of the local energy density $B_i^2$ averaged over the different realizations. This method is accurate when the field is dominated by a single mode. In fig. \ref{fig2}, the time series is displayed for an applied field in the anti-Helmholtz  configuration such that one may expect to measure the decay rate of the quadrupolar mode. The decay of the magnetic energy is not exponential  (black curve) and as will be made clear below, the time recording actually transitions between two different exponential  behaviors. In this case, the decay rate is not correctly measured from the evolution of the magnetic energy. The method that we present now solves this problem. 

We note $\textbf B_1(t)$ (resp. $\textbf B_2(t)$) the vectors defined using the $30$ values measured by the $10$ probes  close to disk $1$ (resp. $2$).  In order to extract the decay of the amplitude of unstable modes, we define two reference geometries  noted $\textbf B_1^{r}$ and  $\textbf B_2^{r}$, which are the time averages of the vectors $\textbf B_1$ and  $\textbf B_2$ when the dynamo is operating. The spatial structure given by  $\textbf B_1^{r}$ and  $\textbf B_2^{r}$ is a dipole, and it does not change significantly above the dynamo onset. Note that below  the onset and for a field  applied in the Helmholtz configuration   the  geometry of the induced magnetic field is quite similar to the dipole geometry $\textbf B_1^{r}$ and  $\textbf B_2^{r}$. Thus we could take indifferently either the spatial structure of the unstable mode or the one of the induced field to measure the decay rates. Therefore, this method can be applied even when the dynamo threshold has not been reached. 
The dipole and quadrupole amplitudes $D(t)$ and $Q(t)$ are defined by
 
\begin{equation}
D(t)=\frac{1}{2}\left( \frac{\textbf B_1 \cdot \textbf B_1^{r}}{ \vert \textbf B_1^r \vert  }+\frac{\textbf B_2 \cdot \textbf B_2^{r}}{\vert \textbf B_2^r \vert  } \right) \quad Q(t)= \frac{1}{2} \left( \frac{\textbf B_1 \cdot \textbf B_1^{r}}{ \vert \textbf B_1^r \vert  }-\frac{\textbf B_2 \cdot \textbf B_2^{r}}{\vert \textbf B_2^r \vert  } \right)
\end{equation}
where $\vert \textbf B_i^r \vert=(\textbf  B_i^r.\textbf B_i^r)^{1/2}$. This amounts to a projection of the measured magnetic field on a reference spatial structure.

\begin{figure}[htb!]
\begin{center}
\includegraphics[width=70mm,height=70mm]{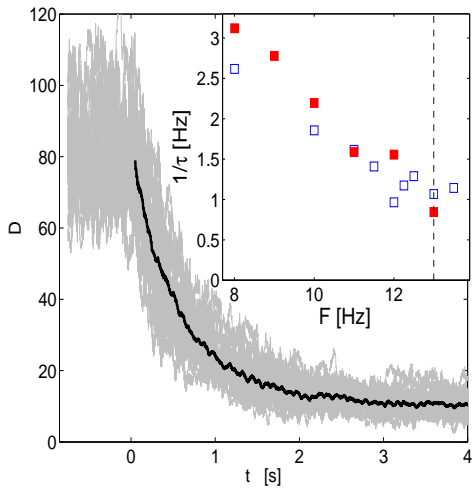}
\caption{Decay of the amplitude of the dipolar projection  for $F=11\,Hz$ (applied field in the Helmholtz configuration). The thick black line is the average over the different realizations (displayed as thin grey lines). Inset: decay rate $1/\tau$ of the dipole as a function of the disk rotation frequency for an applied field in Helmholtz ($\square$) or anti-Helmholtz ($\blacksquare$) configuration. }
\label{fig3}
\end{center}
\vspace{- 1cm}
\end{figure}

We show in fig. \ref{fig3} different realizations of the decay of $D(t)$ (in grey) for  $F=11\,$Hz (the applied magnetic field is shut down at $t=0$). The average over the different  realizations $\bar D(t)$ (black curve) exhibits an exponential decay. $\bar D$ is fitted with an exponential function $ A \exp(-t/\tau)+\bar D_0$, where $\tau^{-1}$ is the decay rate and $\bar D_0$ is the amplitude of the dipole without applied magnetic field.  Indeed, because of the imperfectness of the bifurcation, $\bar D_0$ is not exactly zero. It is created by the different sources of applied field (Earth magnetic field, remanent magnetization of the disks, ...).

\begin{figure}[htb!]
\begin{center}
\includegraphics[width=70mm,height=70mm]{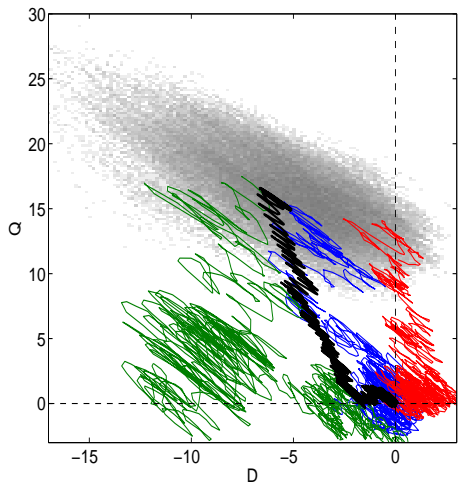}
\caption{Phase-space  for the induced field $(D,Q)$ for $F=11\,Hz$. The grey set contains the trajectories when a field is applied in the anti-Helmholtz configuration. The (colored online, thin) continuous curves are individual trajectories that start when the field is set to zero. The thick black curve is the average over the realizations.}
\label{fig4}
\end{center}
\end{figure}

The measured value of the decay rate is quite robust. As is displayed in the inset of fig. \ref{fig3}, using the projection method, the same results are obtained for the decay rate of the dipole in the Helmholtz (resp. anti-Helmholtz) configuration. We now understand the behavior of the total energy displayed in fig. \ref{fig2} for an anti-Helmholtz applied field. At short time, the quadrupolar component is much larger than the dipolar one. The quadrupolar decay rate being larger than the dipolar one, after an initial phase (here of around $0.2$ s), the energy of the quadrupolar component is drastically reduced and the dipolar component gives the main contribution to the energy. We note that the dipolar and the quadupolar energy add up to nearly  $80$ percent  
of the total energy. All together, this explains why the time recording of the total energy crossovers between two different exponential behaviors.   We also point out that a single configuration of applied field allows to extract both decay rates because the projection disentangles  the relative contribution of each mode. This is mostly achievable with an applied field in the  anti-Helmholtz configuration. Then a dominant part of the energy is initially injected  in the most damped mode so that the slowest decaying mode is ultimately observed even though it has a small initial energy mostly due to imperfection in the experimental set-up or in the symmetry of the applied field. 

Trajectories in phase-space also display the two successive behaviors, as shown in fig. \ref{fig4}. When a field is applied in  the anti-Helmholtz configuration, the trajectories  wander around a wide spot (in grey). Indeed turbulent fluctuations transfer energy between the modes.  When the applied field is set to zero, the trajectories relax toward the origin. Turbulent fluctuations are also responsible for the observed  large variability between the realizations. In contrast, the average over the realizations (thick black line)  is simpler and is made of an evolution toward the $Q=0$ axis followed by the evolution toward $D=Q=0$. Despite the fluctuations,   we thus can observe in phase-space the two successive behaviors (decrease of the quadrupolar component followed by the decrease of the dipolar one).

We compare the decay rates for the dipolar and quadrupolar mode in fig. 5. For disks fitted with curved blades (upper panel), we observe that at onset ($F=13$ Hz), the decay rate of the dipole has been reduced by a factor $2.6$ compared to its value at $F=8$ Hz.  The finite value of the decay rate measured at $F=13$ Hz is probably biased by the imperfection of the bifurcation  (see the discussion in \cite{P10}). We note that the decay rate of the quadrupolar mode has also been reduced from $5 s^{-1}$ to nearly $3 s^{-1}$ at threshold. Within the accuracy of the measurements, the variations of the decay rates are linear in $F$ in the range $8-13$ Hz \cite{notespeed}. Both curves intersect the $1/\tau=0$ line at comparable rotation frequencies close to $17$ and $20$ Hz respectively.

\begin{figure}[htb!]
\begin{center}
\includegraphics[width=70 mm]{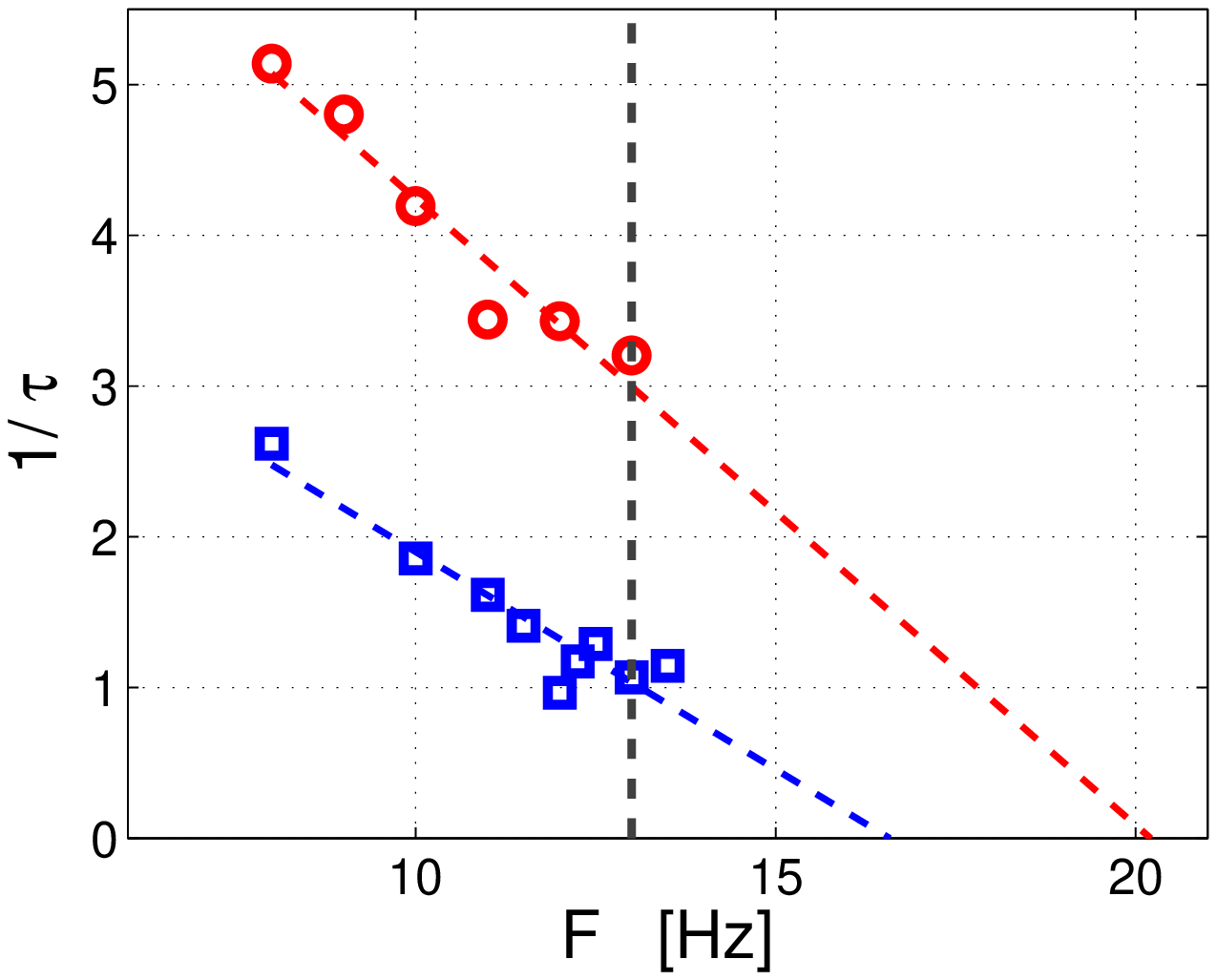}
\includegraphics[width=70 mm]{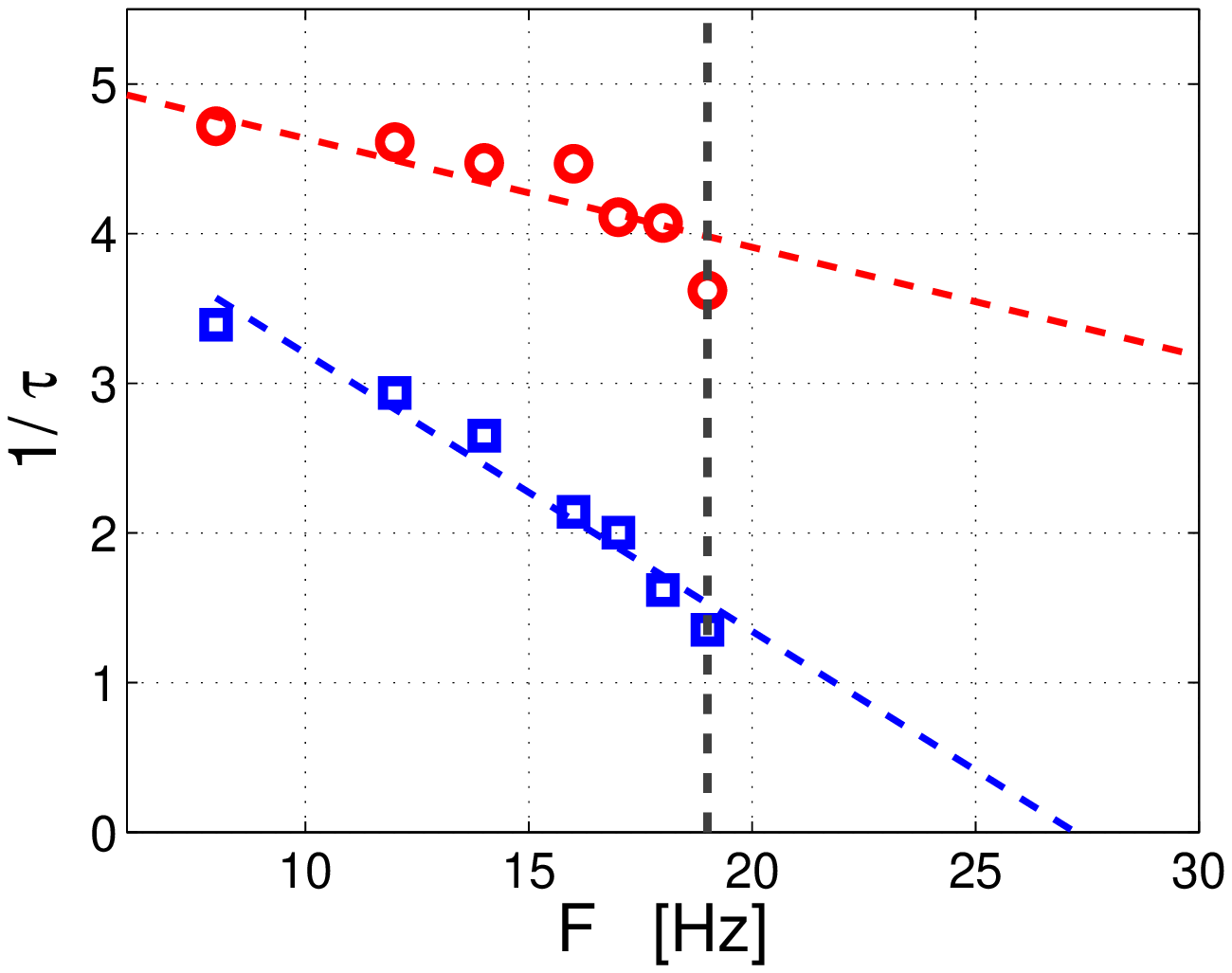}
\end{center}
\caption{Decay rate of the dipolar mode (blue square) and of the quadrupolar mode (red circle) as a function of the rotation frequency. Upper panel: the flow is driven by disks with curved blades. Lower panel: the flow is driven by disks with straight blades.}
\label{fig5}
\end{figure}

A second set of experiments was performed with disks fitted with straight blades. In exact counter-rotation, the onset takes place at a larger value $F=19$ Hz. Between $F=8$ Hz and the onset, the dipolar decay rate is reduced by a factor $2.5$. The amplitude of the variation is similar to the one observed in the case of disks fitted with curved blades. In contrast, the quadrupolar decay rate  only varies from $4.8 s^{-1}$ to $3.6 s^{-1}$. The linear fits of these curves cross the $1/\tau=0$ line at $F=27$ Hz for the dipolar mode and at $F$ larger than $70$ Hz for the quadrupolar one. We note that the estimated decay rates display strong fluctuations close to the dynamo threshold and thus can be sensitive to the fitting parameters. However, it is clear that the quadrupolar decay rate displays a stronger variation with $F$ in the case of curved blades. Thus, at dynamo threshold, the quadrupolar mode is closer to its instability threshold when the flow is driven by curved blades. 

These observations agree with the mechanism proposed to explain the dynamical regimes observed in the experiments \cite{model1}. The dynamical regimes (f.i. random or periodic reversals) take place when both the dipolar and the quadrupolar modes are close to their onset of instability. Then breaking the forcing symmetry, {i.e.} rotating with $F1\neq F2$, couples the two modes that can achieve a saddle-node bifurcation. Above this bifurcation, periodic reversals are observed while close to the threshold of the saddle-node bifurcation, turbulent fluctuations trigger random reversals. Reversals are observed when the effect of the symmetry breaking is comparable to the difference between the growth rates of the two modes. In the case of disks fitted with curved blades, the difference in values $F$ at which $1/\tau$ reaches $0$ is equal to $\Delta F \simeq 3$ Hz. Dynamical regimes of the magnetic field can be observed provided that $\vert F_1 - F_2 \vert$ is larger than roughly $4$ Hz. 

In the case of disks fitted with straight blades, we have concluded from fig. \ref{fig5} that the difference between the growth rates is quite large. We thus expect that the breaking of symmetry will not be efficient enough to generate the saddle-node bifurcation. Indeed, this is confirmed by the experimental results since no dynamical regimes are actually observed with these disks. In other words, the flow driven by these disks generates modes with strongly different  thresholds. Breaking the symmetry by rotating the disks at different speed does not result in a  strong enough coupling between the two modes so that no field reversals can be generated.  

 \begin{figure}[htb!]
\begin{center}
\includegraphics[width=70 mm]{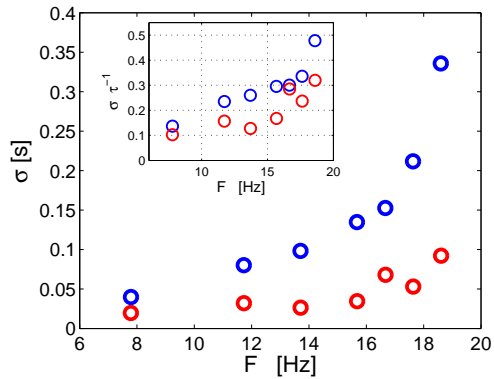}
\end{center}
\caption{Standard deviation of the decay rate of the  quadrupolar (red circle) and dipolar mode (blue square). In the inset,  standard deviation  over the mean.}
\label{fig6}
\end{figure}

Finally, we discuss some statistical properties of the decay rate measured by the projection method. The decay rate $\tau$ varies from one realization to the other, as displayed in fig. 3. 
The standard deviation $\sigma$ of $\tau$  is displayed in fig. 6. It increases in the vicinity of  the onset of the dynamo instability. Values up to $50$ percents are achieved for the fluctuations over the mean, $\sigma \tau^{-1}$. This explains why large samples are required to converge the decay rate close to the dynamo onset.

\acknowledgements
We acknowledge our colleagues of the VKS team with whom the experimental data used here have been obtained.


\begin{thebibliography}{99}
\bibitem{notespeed} Mechanical constraints prohibit long measurements to be performed at rotation frequencies smaller than $8$ Hz.
\end{thebibliography}
\end{document}